\documentclass[12pt]{iopart}
\usepackage[square,numbers]{natbib}

\usepackage{xcolor}
\usepackage{graphicx}
\usepackage{caption}
\usepackage[colorlinks=true,linkcolor=blue,citecolor=blue,urlcolor=blue]{hyperref}
\usepackage{tikz}
\usepackage{braket}
\usepackage[protrusion=true, expansion=true]{microtype}
\begin{document}
\title{Gate teleportation-assisted routing for quantum algorithms}
\author{Aravind Plathanam Babu}
\address{Nano and Molecular Systems Research Unit, University of Oulu, Finland}
\ead{aravind.babu@oulu.fi}
\author{Oskari Kerppo}
\address{Faculty of Information Technology, University of Jyväskylä, Finland\\ Quanscient Oy, Tampere, Finland} 
\author{Andrés Muñoz-Moller}
\address{Faculty of Information Technology, University of Jyväskylä, Finland}
\author{Majid Haghparast}
\address{Faculty of Information Technology, University of Jyväskylä, Finland}
\author{Matti Silveri}
\address{Nano and Molecular Systems Research Unit, University of Oulu, Finland}
\begin{abstract}
The limited qubit connectivity of quantum processors poses a significant challenge in deploying practical algorithms and logical gates, necessitating efficient qubit mapping and routing strategies. When implementing a gate that requires additional connectivity beyond the native connectivity, the qubit state must be moved to a nearby connected qubit to execute the desired gate locally. This is typically achieved using a series of SWAP gates creating a SWAP path. However, routing methods relying on SWAP gates often lead to increased circuit depth and gate count, motivating the need for alternative approaches.
This work explores the potential of teleported gates to improve qubit routing efficiency, focusing on implementation within specific hardware topologies and benchmark quantum algorithms. We propose a routing method that is assisted by gate teleportation. It establishes additional connectivity using gate teleportation paths through available unused qubits, termed auxiliary qubits, within the topology. To optimize this approach, we have developed an algorithm to identify the best gate teleportation connections, considering their potential to reduce the depth of the circuit and address possible errors that may arise from the teleportation paths. Finally, we demonstrate depth reduction with gate teleportation-assisted routing in various benchmark
algorithms, including case studies on the compilation of the Deutsch-Jozsa algorithm
and the Quantum Approximation Optimization Algorithm (QAOA) for heavy-hexagon
topology used in IBM 127-qubit Eagle r3 processors. Our benchmark results show a $10-25\%$ depth reduction in the routing of selected algorithms compared to regular routing without using teleported gates.
\end{abstract}
\maketitle
\section{Introduction}
Executing a quantum algorithm on the quantum computer requires a transpilation process, translating the quantum circuit describing the algorithm into machine-readable instructions tailored to the hardware. This process ensures compatibility with the native gate set, qubit connectivity, and other hardware constraints while optimizing performance metrics such as circuit depth and number of gates. An essential step in the transpilation process is mapping and routing. Mapping involves assigning logical qubits from the quantum algorithm to the physical qubits available on the quantum computer. Routing refers to modifying quantum circuits to ensure that they adhere to the connectivity constraints of the target quantum computer.

The qubit connectivity varies among different hardware platforms. Some hardware platforms, such as trapped-ion- or neutral-atom-based quantum computers~\cite{Trappedion1, Trappedion2,atom-array}, offer greater flexibility to reconfigure connectivity, facilitating effective all-to-all interactions. In contrast, superconducting quantum computers typically exhibit a fixed topology, where each qubit interacts only with a limited number of neighbouring qubits~\cite{supercondu, supercondu1}. The implementation of two-qubit gates between qubits that are not directly connected is typically achieved by exchanging the state of the qubits within the circuit through a series of SWAP gates. 
Although this method is simple, it considerably increases the circuit depth and the total number of gates required. This can lead to an increase in gate errors, which ultimately results in decreased performance.

Most of the current routing methods combine mapping and SWAP-based routing schemes. These methods use classical optimization algorithms to map logical qubits of quantum algorithms to physical qubits on quantum hardware and to determine the optimal placement of SWAP gates for routing. These problems can be formulated using graph theory: the mapping problem resembles subgraph isomorphism, while the routing problem is closely related to the token-swapping problem~\cite{cowtan2019qubit}. As the token-swapping problem is classified as NP-hard, the routing problem is believed to belong to the same complexity class~\cite{cowtan2019qubit, PRXQuantumAdvantagesR}. Current routing algorithms are based on various classical techniques, including shortest-path search~\cite{saeedi2011synthesis,shortestpath2}, heuristic methods~\cite{Sabre2019, Lao2021, Li2020,Liu2022, Nishio2020}, or other optimization techniques \cite{chakrabarti2011linear,shafaei2013optimization,wille2014optimal,lye2015determining,bhattacharjee2017depth,venturelli2017temporal,venturelli2018compiling,booth2018comparing,oddi2018greedy}. However, these algorithms suffer from scalability issues and generally produce a suboptimal solution with high SWAP overhead. Recently developed reinforcement learning-based routing methods have demonstrated better scalability and the potential to improve routing efficiency~\cite{tang2024alpharouter, RL_based_David}. Software development frameworks such as Qiskit use the SWAP-based bidirectional heuristic search algorithm (SABRE)~\cite{Sabre2019} and its variants. The t$\ket{\rm{ket}}$ framework~\cite{Sivarajah2020} also employs heuristic methods and iterative graph-based approaches to improve performance and ensure compatibility with various quantum hardware architectures.

Quantum state teleportation~\cite{Bennett1993} is a protocol in quantum information theory without a classical analogue that allows one to move a quantum state between two parties over an arbitrary distance using entangled states and classical communication. Similarly, quantum gate teleportation~\cite{Gottesman1999} is a protocol that implements quantum gates between spatially separated qubits using shared entanglement in auxiliary qubits and classical communication. Quantum state and gate teleportation have been shown to have applications for quantum communication tasks, such as quantum networks~\cite{Gisin2007}, quantum repeaters~\cite{Briegel1998}, and quantum cryptography~\cite{Gordon2010}. In quantum computing, quantum teleportation is necessary for measurement-based quantum computing~\cite{Raussendorf2001}, and also serves as a primitive for universal fault-tolerant quantum computing~\cite{Gottesman1999}.

Recent advancements in superconducting hardware have introduced the possibility of dynamic circuits, which utilize mid-circuit measurements and classically controlled gates to expand the range of operations accessible to quantum computers. In Ref.~\cite{Bumer2024}, experimental implementations of dynamic circuits have been demonstrated, notably performing a CNOT gate between distant qubits and preparing large multipartite entangled states. Subsequently similar techniques were used to perform experimental $n$-qubit Fourier transform using dynamic circuits~\cite{Baumer2024_2}. Both articles show that mid-circuit measurements and feed-forward operations can be used to reduce the depth of circuits, which may decrease the effects of decoherence and increase the fidelity of quantum algorithms.

Qubit routing methods typically move around the state of a qubit using SWAP operations. Similarly, quantum teleportation can be used to teleport the state of a qubit to an arbitrary position within the topology of the quantum computer, as long as it is possible to connect the starting position to the final position using entangled states through auxiliary qubits and perform classically controlled operations upon measuring those auxiliary qubits. An in-depth analysis of the speed-up of the local operations and classical communication (LOCC) assisted routing technique over SWAP-based methods is studied in Ref.~\cite{Devulapalli2024}, as long as $\mathcal{O}(n)$ auxiliary qubits are allowed. In Ref.~\cite{Hillmich2021}, quantum teleportation is proposed as a complementary method in routing, where the possible teleportation channels through ancillary qubits add virtual edges to the coupling map of a given quantum hardware topology, increasing the possible paths available to the A$^*$-based mapping algorithm~\cite{Zulehner2019}, thus achieving a lower circuit depth. In Ref.~\cite{Padda2024}, a quantum routing method with state teleportation is proposed using entanglement generating devices such as photonic interconnects, creating the required channel for teleportation by entangling qubits with photons, and then performing Bell measurements to swap the entanglement from a pair of qubit-photon entangled states to a qubit-qubit EPR pair, without the need for additional ancillary qubits and gates within the QPU.


In a similar way to quantum state teleportation, quantum gate teleportation can be used to implement a gate between the qubits that are not directly connected over arbitrary distances within a quantum processor topology. The implementation of the gate teleportation scheme on a given hardware involves establishing a virtual connection between the control and target qubits using auxiliary qubits. These auxiliary qubits can be selected from the unused qubits in the hardware after mapping all data qubits. For example, if we want to run a 20-qubit algorithm on a 50-qubit device, we will have 30 auxiliary qubits available to establish the teleportation path. Teleporting all non-local gates may not be optimal, as it can increase circuit complexity and accumulate errors. Therefore, a key challenge is to identify which gates should be teleported to achieve optimal routing. We address this critical question of identifying which gates need to be teleported and determining where teleportation gates should be introduced. To this end, we have developed the Routing with Teleported Gates (RTG) method, a systematic approach designed to identify optimal virtual connections and routing pathways utilizing these connections. We introduce gate teleportation as an add-on to existing SWAP-based routing and mapping methods. This hybrid approach selectively employs teleportation for specific gates, balancing the trade-off between circuit depth, gate count, and error, thereby enhancing the overall efficiency of quantum circuit execution.  We quantify the results by examining the depth and number of two-qubit gates required for various algorithms. Finally, we present a case study of the routing of the Deutsch-Jozsa and the Quantum Approximation Optimization Algorithm (QAOA) on the heavy-hexagon topology of IBM 127-qubit Eagle r3 processors.

This work is structured as follows. Section~\ref{Sec.Methods} presents the circuit-level implementation of gate teleportation, followed by the routing with teleported gates (RTG) algorithm and the noise-aware RTG method. Section~\ref{Sec.CNOTs} discusses results for benchmark algorithms, quantifying circuit depth and the number of two-qubit gates required. Finally, Sections~\ref{sec.DJ}-\ref{sec.QAOA} provides case studies on routing the Deutsch-Jozsa and QAOA for the IBM 127-qubit Eagle r3 processor topology.
\section{Methods}\label{Sec.Methods}
In this section, we describe our routing protocol that employs quantum gate teleportation specifically for CNOT gates and controlled single-qubit unitary gates. An efficient method for the CNOT gate teleportation is already present in Ref.~\cite{Bumer2024}, and it can be extended to the case of teleporting an arbitrary controlled unitary (CU) gate, as the optimal local implementation of controlled single qubit unitary gates is well-known, as shown in Ref.~\cite{Eisert2000}.

\begin{figure}[h!]
\includegraphics[width=1\linewidth]{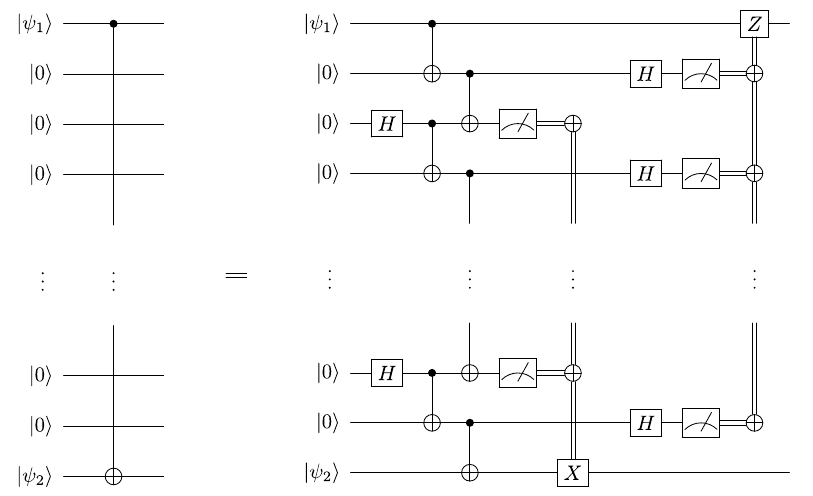}
   \caption{Circuit diagram schematic of a gate-teleportation implementation of a long-distance CNOT by utilizing mid-circuit measurements and feed-forward operations}
  \label{fig:cnot}
\end{figure}

\begin{figure}[h!]
\includegraphics[width=1\linewidth]{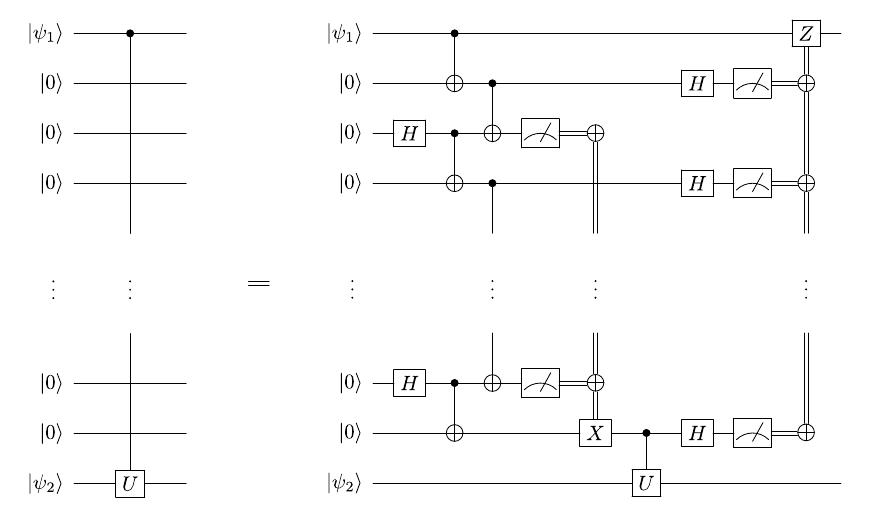}
     \caption{Circuit diagram schematic of a gate-teleportation implementation of a long-distance controlled unitary gate (CU) by utilizing mid-circuit measurements and feed-forward operations}
  \label{fig:cu}
\end{figure}

The teleportation of controlled gates requires both the target and control qubits to be locally connected with pre-entangled auxiliary qubits in a 1D chain topology or at least connected to their nearest neighbours. Teleportation protocol consists of a series of measurements on the auxiliary qubits, followed by conditional operations applied to the target and control qubits based on the outcomes of these measurements. Teleporting a CNOT gate through an auxiliary path with $N$ qubits requires $N+1$ CNOT gate operations and $N$ mid-circuit measurements. However, only two of these CNOT gates operate directly on the data qubits—one acts on the control qubit and the other acts on the target qubit. Similarly, controlled $U$ teleportation requires $N$ CNOT gate operations, one controlled $U$ gate, and $N$ mid-circuit measurements. Importantly, the depth of these teleportation circuits remains constant regardless of the distance~$N$. The circuits for the CNOT and the controlled $U$ teleportations over arbitrary distances are illustrated in Figs.~\ref{fig:cnot} and \ref{fig:cu}.


\begin{figure}[t!]
\includegraphics[width=\linewidth]{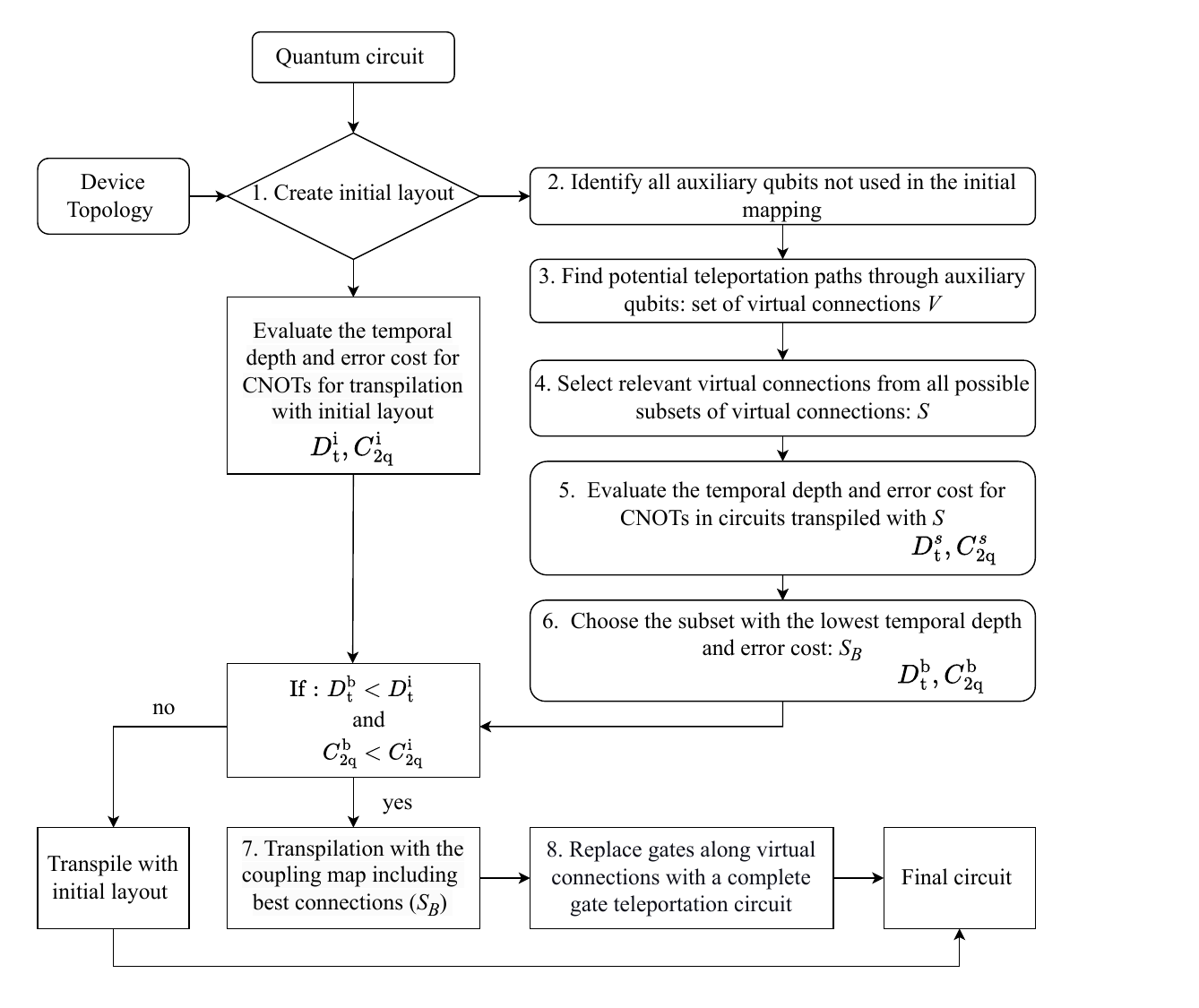}\caption{Flowchart representing the implementation of noise-aware Routing with Teleported Gates (RTG) method. Here, \( D_{t}^{x} \) and \( C_{\rm 2q}^{x} \), with \( x \in \{{\rm i}, s, {\rm b}\} \), represent the temporal depth and error cost for routing under different conditions:  
\( x = {\rm i} \) corresponds to the initial layout,  
\( x = s \) refers to routing with a subset \( S \) of possible virtual edges, and  
\( x = b \) denotes the best subset of connection that minimizes both depth and error cost. The RTG method follows the same steps but does not incorporate error cost in the optimization.}
\label{fig:flowchart}
\end{figure}

We now describe how to identify and utilize gate teleportation connections in a given topology for the circuit corresponding to a quantum algorithm using our Routing with Teleported Gates (RTG) method and the noise-aware version of RTG. The complete process of the noise-aware RTG method is illustrated in Fig.~\ref{fig:flowchart}.  The RTG method follows the same steps but does not incorporate noise considerations.
RTG method starts with mapping logical qubits to physical qubits within the given topology using existing mapping methods, such as SABRE~\cite{Sabre2019}. Generally, after this step,  the coupling map that describes the native connectivity for the initial layout is utilized for subsequent routing procedures. The next step involves identifying unused qubits, referred to as auxiliary qubits, and exploring potential paths that can be established using these auxiliary qubits for teleportation. These paths, known as virtual edges, allow data qubits to interact indirectly via the auxiliary qubits, effectively extending the connectivity of the hardware beyond its native coupling map. In the subsequent and most crucial process (steps 4 to 6 of Fig.~\ref{fig:flowchart}), the method evaluates which virtual edges would improve the routing compared to standard routing on the native coupling map. However, virtual edges differ from native physical connections in two key aspects: the gate times are longer, and the gate errors are higher than those of native two qubit gates. Therefore, these factors must be carefully considered when evaluating the routing improvement.

Typically, circuit depth or the number of gates is used as a metric to quantify routing improvement. However, circuit depth only measures the number of sequential “layers” of quantum gates executed in parallel and does not account for the actual runtime of these layers. For example, a layer containing a teleported two-qubit gate will take longer to execute compared to a regular layer with native two-qubit gates due to the additional overhead of teleportation. In the RTG method, we consider temporal depth to evaluate the runtime differences caused by variations in gate execution times. The temporal depth is defined as the maximum cumulative gate execution time along any path from the start to the end of the circuit. The temporal depth is estimated from the Directed Acyclic Graph (DAG)~\cite{Sabre2019} representation of the circuit using the following formula
\begin{equation}
    D_{\rm t} = t_g^{\rm 1q} N_L^{\rm 1q} + t_g^{\rm 2q} N_L^{\rm 2q} + t_g^{\rm tele} N_L^{\rm tele},
\end{equation}
where $N_L^{\rm 1q}$ is the number of layers containing only single qubit gates, $N_L^{\rm 2q}$ is the number of layers containing at least one native two-qubit gate and no teleported gates, and $N_L^{\rm tele}$ is the number of layers containing at least one teleported two-qubit gate. The parameters $t_g^{\rm 1q}$, $t_g^{\rm 2q}$, and $t_g^{\rm tele}$ represent the execution times for single qubit gates, native two qubit gates, and teleported two qubit gates, respectively.

The RTG searches through the available virtual connections to select the best ones that can reduce temporal depth compared to standard transpilation without virtual edges. While it is theoretically possible to perform an exact or exhaustive search to evaluate all possible virtual connections, this approach quickly becomes impractical as the number of connections increases. The computational resources required to explore every potential connection grow exponentially with the number of connections and the size of the circuit, making it infeasible for circuits with many auxiliary qubits. Therefore, we develop and benchmark a heuristic search algorithm, which takes into account the connectivity requirements of the circuit. 
The algorithm primarily employs a greedy heuristic combined with randomized transpilation with the SABRE routing method. It begins by preprocessing the circuit (more accurately, the circuit that corresponds to the algorithm) to identify pairs of qubits that require connectivity beyond what is available in the initial layout, using a directed acyclic graph (DAG) representation. From the set of all available virtual connections, the algorithm selects a subset that is either directly required or spatially close to the required qubit pairs (step 4 in Fig. \ref{fig:flowchart}). Another condition considered is that multiple implementations of the teleported gate may not always yield better performance, given the increased error rates and the cost of reset operations required on the auxiliary qubits after every teleported gate operation through the same path. As a result, the search can be constrained to only consider connections that are not used repeatedly. For example, a limit can be set such that only connections repeated twice or less in the whole algorithm are considered. 
Following this, the algorithm generates all valid combinations of subsets of virtual connections, ensuring that no two connections in a subset share the same qubits. This pruning strategy helps filter out combinations that reuse the same set of qubits for teleportation. For each subset of virtual connections, the algorithm computes the cost (temporal depth) and selects the one with the lowest value. The subset that minimizes the cost is stored as the best solution. Multiple transpilation trials are performed for each subset using different random seeds to find the lowest cost. The randomized search approach mitigates the risk of local minima. At the implementation level, the circuit is first transpiled with the best virtual connections identified by the RTG method (step 8. of Fig.~\ref{fig:flowchart}). These virtual connections are then replaced with the corresponding physical implementation teleportation circuits, enabling efficient routing while adhering to hardware constraints.

To further enhance our RTG method, we introduce a noise-aware version of the algorithm, which incorporates the error costs of the teleported gates. This approach recognizes that minimizing temporal depth alone may not always lead to the most reliable circuit execution, as gate errors also significantly impact overall circuit fidelity. In the noise-aware RTG method, the error cost $C_{\rm 2q}$ for all the two-qubit gates of the algorithm is estimated with the formula \cite{errorestimation}: 
\begin{equation}
    C_{\rm 2q}=1-(1-P_{\rm g}^{\rm 2q})^{N_{\rm g}}(1-P_{\rm g}^{\rm tele})^{N_{\rm tele}},
\end{equation}
where $P_{\rm g}^{\rm 2q}$ and $P_{\rm g}^{\rm tele}$ are the average error probabilities for native two-qubit gate and teleported two-qubit gates, respectively. The number of native two-qubit gates is denoted by $N_{\rm g}$ and the number of teleported two-qubit gates by $N_{\rm tele}$.  The two-qubit error cost estimates the cumulative probability of errors by considering the likelihood of errors during both native and teleported two-qubit operations. The noise-aware RTG method selects connections with lower temporal depth and reduced error costs compared to standard transpilation without virtual edges. For each subset of virtual connections, the algorithm computes the temporal depth and the error cost and selects the one with the lowest values for both. The subset that minimizes both costs is stored as the best solution.  However, it is also possible to assign different weights to these criteria, allowing for a more flexible optimization that balances depth reduction and error budget based on specific hardware characteristics. 

%

\section{Result} \label{sec.results}
\subsection{Benchmark with IBM 127-qubit Eagle r3 processor topology} \label{Sec.CNOTs}
\begin{figure}[ht!]
    \centering
\includegraphics[width=\linewidth]{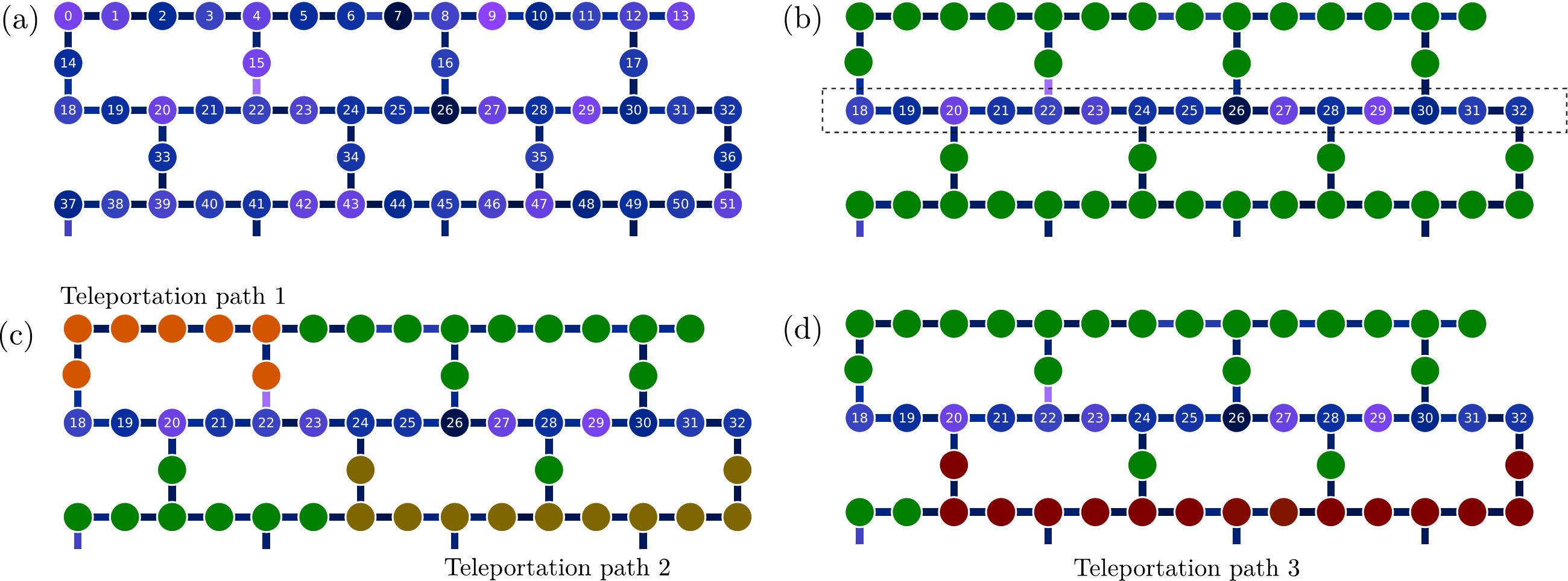}
    \caption{(a) A section of the IBM 127-qubit Eagle r3 processor topology \cite{IBMQuantum}. (b) The initial line layout, with auxiliary qubits highlighted in green. (c) and (d) Examples of possible teleportation paths utilizing auxiliary qubits.}
    \label{IBM_mapping}
\end{figure}

To benchmark the RTG and noise-aware RTG methods, we assess their performance on a segment of the IBM 127-qubit Eagle r3 processor topology, as depicted in Fig.~\ref{IBM_mapping}(a). This device is particularly well-suited for our study, as it supports essential features such as mid-circuit measurements and feed-forward operations. For this analysis, we employ 15-qubit benchmark circuits from the Munich Quantum Toolkit Benchmark Library (MQT Bench)~\cite{quetschlich2023mqtbench}, which encompasses a variety of quantum algorithms. The initial qubit layout is chosen as a simple line topology [qubits no. 18–32 in Fig.~\ref{IBM_mapping}(a)], allowing for a clear demonstration of teleportation paths. While this particular layout is a trivial choice, alternative mapping strategies can be used equally with the RTG method. Following this, auxiliary qubits, indicated in green in Fig.~\ref{IBM_mapping}(b), are identified to find potential teleportation paths. Some possible teleportation paths and their associated auxiliary qubits are illustrated in Fig.~\ref{IBM_mapping}(c-d). It is worth noting that although only three paths are presented for illustrative purposes, many other similar configurations are possible within this topology.

\begin{figure}[ht!]
    \centering
\includegraphics[width=\linewidth]{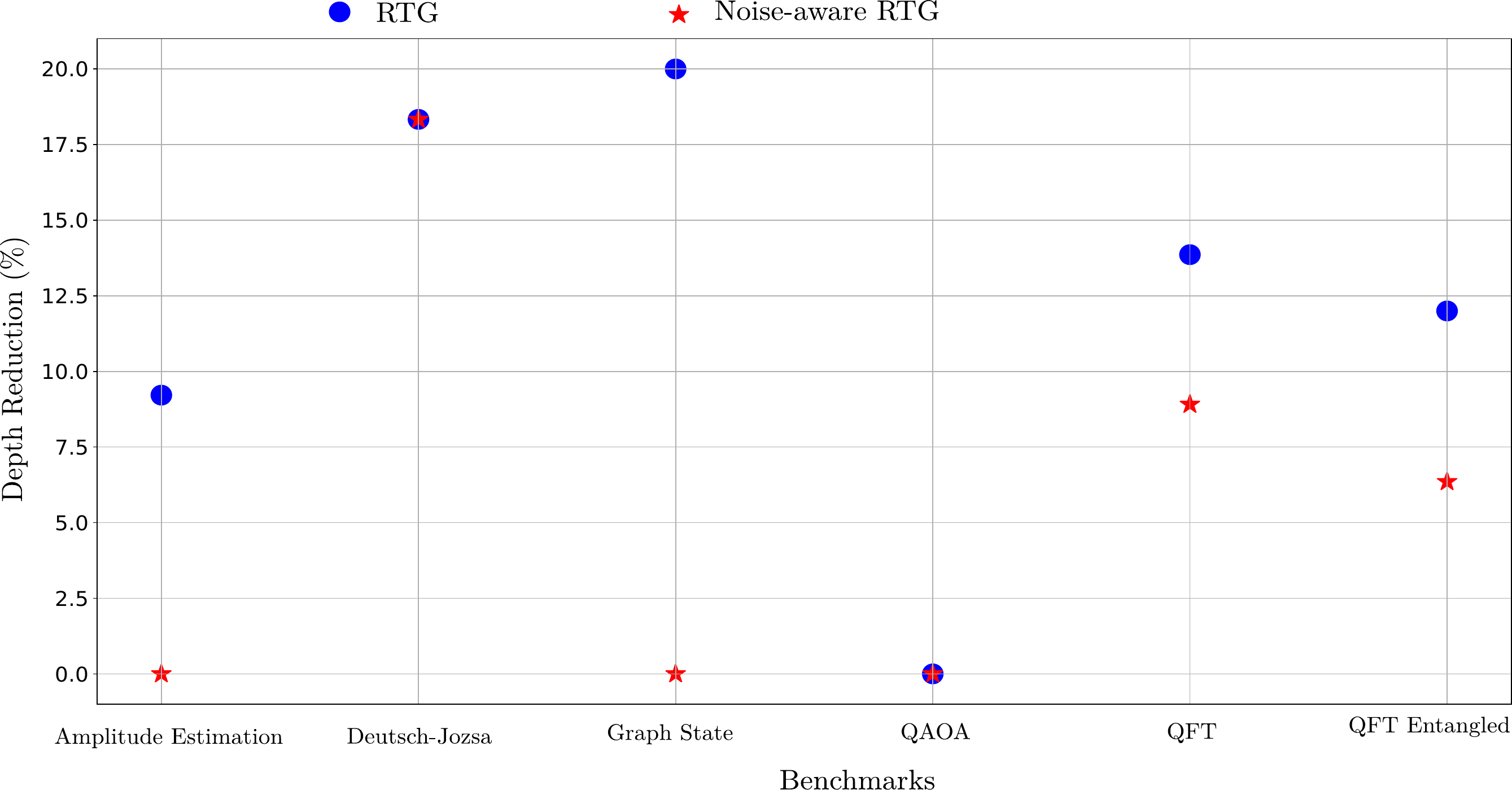}
    \caption{The percentage of depth reduction achieved compared to the initial line topology for routing with RTG and noise-aware RTG methods for various 15 qubit benchmark algorithms. The model topology is based on the IBM 127-qubit Eagle r3 processor architecture.}
    \label{Results1}
\end{figure}

In Fig.~\ref{Results1}, we present the depth reduction achieved using the RTG and noise-aware RTG methods compared to the initial line map topology with SABRE routing. For this analysis, we consider the CNOT gate teleportation and assume the gate time for teleported  CNOT gates is $t_{\rm g}^{\rm tele}=3t_{\rm g}^{\rm 2q}$ and average error probability $P_{\rm g}^{\rm tele}=10P_{\rm g}^{\rm 2q}$. It is important to note that current devices exhibit significantly higher error probabilities; for instance, as reported in Ref.~\cite{Bumer2024},  $P_{\rm g}^{\rm tele}\approx 100P_{\rm g}^{\rm 2q}$, for a distance of teleportation over 10 qubits. As these values are obtained from the first generation of processors capable for mid-circuit measurements and feed-forward, we chose to use values that we expect to be achievable in future devices through improvements in feed-forward times and gate fidelities.  We consider the following 15-qubit benchmark circuits for amplitude estimation: Deutsch-Jozsa algorithm, graph state preparation, quantum approximation optimization algorithm (QAOA), quantum Fourier transform (QFT), and QFT applied to entangled qubits (QFT entangled).  The results demonstrate a reduction in circuit depth across all cases except for QAOA. The Deutsch-Jozsa algorithm achieves approximately $ 18 \%$ reduction in depth when utilizing both RTG and noise-aware RTG methods.  However, performance decreases for the graph state preparation and QFT benchmarks when noise-aware routing is considered (noise-aware RTG). In the case of QAOA, the benchmark circuit requires multiple parametric two-qubit  $Z \otimes Z $ interactions (i.e., $\rm{RZZGate}$), which are then decomposed into two CNOT gates separated by a single qubit Z rotation. As a result, each connection involves more than two CNOT gates, and the heuristic search is constrained to avoid repeated use of the same connections, considering the error of repeated teleportation through the same path. This suggests that teleporting RZZ gates may be a more effective option than teleporting CNOT gates directly in this case. This is indeed the case, see Sec.~\ref{sec.QAOA}, where we demonstrate that the teleportation of controlled-U gates yields better results in the context of QAOA. These findings highlight that the effectiveness of the RTG method is highly dependent on the structure of an algorithm and the specific gate operations involved. The RTG method discussed here focuses mainly on teleporting two-qubit gates. However, the framework can be extended to incorporate other teleported gates, such as Toffoli and fanout gates, or to enable constant-depth execution methods \cite{Buhrman2024statepreparation, yeo2025reducing} for quantum circuits for specific applications.

\subsection{Case study 1: Deutsch-Jozsa algorithm with CNOT gate teleportation }\label{sec.DJ}
In Sec. \ref{Sec.CNOTs}, we have demonstrated depth reduction by incorporating optimal virtual edges into the coupling map. However, we have not yet included the actual teleportation circuits for these connections (step 8. of Fig.~\ref{fig:flowchart}). Therefore, the observed depth reduction represents the maximum possible improvement one can expect with virtual edges alone. Here, we present the complete routing process, including the full teleportation circuits utilizing auxiliary qubits, for the Deutsch-Jozsa algorithm on the IBM topology. The benchmark circuit from Ref.~\cite{quetschlich2023mqtbench} is utilized to evaluate the impact of virtual edges on transpilation outcomes at the implementation level. As a baseline, the circuit is first transpiled with $9$–$15$ qubits arranged in a line topology. After this, we incorporate the optimal virtual edges identified by the RTG method for CNOT gate teleportation into the circuit.
We then compare the depth and number of CNOT gates in the baseline line topology with those in the physically routed implementation that includes the virtual edges. The “Virtual c. map” represents the case where possible additional connections are added to the initial coupling map without considering the actual implementation of virtual connections. This situation represents an ideal case, demonstrating the maximum reduction in circuit depth that can be achieved with a specific set of virtual connections. To ensure a fair comparison, we also consider the number of CNOT gates executed between data qubits in the teleportation-based implementation, as well as the number of teleported CNOT gates. The results are presented in Table \ref{DJ-results}.

\begin{table}[h!]
\resizebox{\textwidth}{!}{
\begin{tabular}{|c|cc|cc|cccc|cc|}
\hline
\textbf{Qubits} & \multicolumn{2}{c|}{\textbf{Line c. map }} & \multicolumn{2}{c|}{\textbf{Virtual c. map}} & \multicolumn{4}{c|}{\textbf{Virtual c. map (Impl.)}}   &\multicolumn{2}{c|}{ \textbf{Depth reduction ($\%$)}} \\ \hline
 & Depth & $N_{\rm CNOT}$ & Depth & $N_{\rm CNOT}$ & Depth & $N_{\rm CNOT}$ & $N_{\rm tele}$ & $N_{\rm CNOT}^{\rm data }$   & Expected  &  Impl.\\ \hline
9  & 35 & 20 & 31 & 21 & 31 & 28 & 1 & 20 & 11 & 11 \\ \hline
10 & 42 & 22 & 37 & 19 & 38 & 26 & 1 & 18 & 12 & 10  \\ \hline
11 & 51 & 22 & 38 & 19 & 43 & 26 & 2 & 24 & 25 & 16 \\ \hline
12 & 53 & 20 & 35 & 26 & 43 & 40 & 2 & 24 & 34 & 19 \\ \hline
13 & 59 & 30 & 46 & 27 & 52 & 49 & 2 & 25 & 22 & 12 \\ \hline
 14 &	62&	24&	44&	34	&52	&48	&2	&32	&29 &16\\ \hline 
 15	&67	&26	&57	&35	&62	&61	&2	&33	&15&	7\\ \hline 	
\end{tabular}}
\caption{\textbf{Benchmarking results for implementation of Deutsch-Jozsa algorithm.} We present the depth and number of CNOT gates ($N_{\rm CNOT}$) for three configurations: the line coupling map (line c. map), the line map with virtual edges (virtual c. map), and the actual implementation of the line map with virtual edges (virtual c. map (Impl.)). Here, $N_{\rm tele}$ refers to the number of teleported gates, while $N_{\rm CNOT}^{\rm data}$ indicates the number of CNOT gates applied to the data qubits. In the last column, we display the depth reduction for the line map with virtual edges (Expected) and its actual implementation (Impl.) compared to the line map. Note that in all of the transpilations,  we use the SABRE as the base method for routing. }
\label{DJ-results}
\end{table}
The results demonstrate a reduction in circuit depth compared to the corresponding line coupling map; however, this reduction is less than anticipated with an ideal coupling map incorporating virtual connections (Virtual c. map). This discrepancy occurs because the ideal coupling map regards virtual connections as equivalent to native connections without accounting for physical implementation, leading to an overestimation of depth reduction. In the case of physical implementation, the total number of CNOT gates is considerably higher than in the line topology. Nonetheless, the number of CNOT gates directly acting on data qubits remains relatively similar. The additional increase is attributed to the auxiliary qubits used for teleportation, which necessitate extra CNOT gates to establish the teleportation paths, see Fig.~\ref{fig:cnot}. Reducing circuit depth and the number of gates on data qubits is crucial for optimizing quantum algorithms on near-term devices, where limited coherence times and gate fidelities constrain performance. A shallower circuit helps to reduce decoherence, resulting in improved overall fidelity. Therefore, the reduction in circuit depth discussed here is highly relevant for near-term quantum devices.

\subsection{Case study 2: QAOA algorithm with controlled-U gate teleportation } \label{sec.QAOA}
In the second case study, we consider a QAOA Ansatz circuit for the Max-Cut problem using 9 to 15 qubits. As previously discussed, CNOT gate teleportation is not ideal for QAOA due to the exclusive use of RZZ gates in the Ansatz circuit, which are better candidates for controlled-U teleportation. We consider the model topology of the IBM 127-qubit Eagle r3 processor architecture. The initial mapping is a line coupling map of qubits $18$–$32$. We further evaluate the routing improvements achieved by adding two virtual connections between qubits $20$-$28$ and $22$-$30$. These virtual connections are physically routed through auxiliary qubit paths connecting the respective qubits, see Fig.~\ref{fig:cu}. After the virtual RZZ gates are implemented, all other gates are decomposed into CNOTs to enable a more relevant comparison. The results are presented in Table \ref{QAOA-results}.

\begin{table}[h!]
\resizebox{\textwidth}{!}{
\begin{tabular}{|c|cc|cc|cccc|cc|}
\hline
\textbf{Qubits} & \multicolumn{2}{c|}{\textbf{Line c. map }} & \multicolumn{2}{c|}{\textbf{Virtual c. map}} & \multicolumn{4}{c|}{\textbf{Virtual c. map (Impl.)}}   &\multicolumn{2}{c|}{ \textbf{Depth reduction ($\%$)}} \\ \hline
 & Depth & $N_{\rm CNOT}$ & Depth & $N_{\rm CNOT}$ & Depth & $N_{\rm CNOT}$ & $N_{\rm tele}$ & $N_{\rm CNOT}^{\rm data }$   & Expected  &  Impl.\\ \hline
9  & 130 & 107 & 108 &77 &116 & 184 & 13 & 60 & 17 & 11\\ \hline
10 & 150 & 129 & 115 &81 & 144 & 229 & 14 & 64 & 23 &  4 \\ \hline
11 & 179 & 180 &  108 & 115&134 & 159 & 6  & 101 & 40 & 25 \\ \hline
12 & 182 & 193 & 130 &125 &147 & 169 & 6  & 107 & 29 & 19 \\ \hline
13 & 183 & 184 & 127 &129 &167 & 333 & 23 & 96 & 31 & 9 \\ \hline
14 & 186 & 201 & 139 &129 &162 & 215 & 11 & 103 & 25 & 13 \\ \hline
15 & 209 & 205 & 139 &113 &159 & 269 & 13 & 118 & 33 & 24 \\ \hline	
\end{tabular}}
\caption{\textbf{Benchmarking results for implementation of QAOA algorithm with controlled-U teleportation}. We present the depth and number of CNOT gates ($N_{\rm CNOT}$) for three configurations: the line coupling map (line c. map), the line map with virtual edges (virtual c. map), and the actual implementation of the line map with virtual edges [Virtual c. map (Impl.)]. Here, $N_{\rm tele}$ refers to the number of teleported gates, while $N_{\rm CNOT}^{\rm data}$ indicates the number of CNOT gates applied to the data qubits. In the last column, we display the depth reduction for the line map with virtual edges (Expected) and its actual implementation (Impl.) compared to the line map.}
\label{QAOA-results}
\end{table}
The results again demonstrate depth reduction with teleported gates; however, the extent of the reduction is highly dependent on the number of qubits. This issue arises because the connections were not optimized for the specific number of qubits; instead, two fixed connections were used in all instances. The total number of CNOT gates in the circuit is relatively high, but the number of CNOTs acting on the data qubits is comparable to that of the line map. The number of CNOTs acting on the data qubits is a critical factor, as the auxiliary qubits are used solely for teleportation and are discarded after mid-circuit measurements. Furthermore, it may be possible to implement error mitigation techniques during these measurements to further enhance performance~\cite{Cai23}.  It is also important that the number of teleported gates is relatively high in this scenario, as no constraints were introduced, unlike in the previous case. Consequently, while the results show potential, they may not translate into optimal performance on current quantum devices due to these limitations. 
\section{Conclusions} \label{sec.conc}
Efficient qubit mapping and routing are critical for implementing practical quantum algorithms, particularly as the hardware constraints, such as limited qubit connectivity of current devices, significantly impact circuit performance. In this work, we explored teleportation-assisted routing strategies to improve the routing for the quantum algorithms. We proposed the RTG (Routing with Teleported Gates) method, which introduces virtual connections through gate teleportation paths by utilizing available auxiliary qubits within the topology. By designing a heuristic algorithm, we identified optimal teleportation paths that balance the trade-offs between circuit depth reduction and potential errors arising from gate teleportation. 

Our evaluation of routing performance on benchmark circuits—including the Deutsch-Jozsa algorithm, graph state, QFT, and QAOA—demonstrated that teleportation-assisted routing can reduce circuit depth by $10$ to $20 \%$ compared to a native line topology. While teleporting CNOT gates was generally effective, our case study on QAOA circuits showed that teleporting RZZ gates using controlled-U teleportation provided better performance. These findings highlight the potential of teleported gates to improve routing efficiency and emphasize the importance of customizing the routing strategy to align with specific gate requirements and device topology. Despite these promising results, certain limitations were identified. The overhead introduced by auxiliary qubits during teleportation must be carefully managed to ensure the practicality of this method for current quantum devices. While the total number of CNOT gates in the circuit is relatively high, the number of CNOTs acting on the data qubits remains comparable to that of a line topology. This is particularly relevant since the auxiliary qubits are solely used for teleportation and are discarded after mid-circuit measurements. The scaling of auxiliary qubit usage depends on algorithmic requirements and the number of available qubits on the hardware, making it an adaptable strategy rather than a fixed overhead. Additionally, incorporating error mitigation techniques for these measurements could further enhance performance~\cite{Cai23}, making teleportation-based routing more viable on noisy intermediate-scale quantum (NISQ) devices. Note that the choice of the initial layout could also affect the results, and the RTG method could utilize existing routing and mapping techniques to optimize the initial layout based on the given algorithm and hardware topology. In a more advanced routing scheme, potential teleportation paths could also be considered during initial layout selection to further enhance efficiency. Alternatively, a dedicated subset of qubits within the hardware could be specifically assigned for teleportation, ensuring that idle qubits are optimally positioned to facilitate nonlocal operations. 

Overall, this work demonstrates the feasibility of implementing teleportation-assisted routing as a powerful tool for enhancing qubit connectivity in near-term quantum devices. By further optimizing teleportation paths and reducing the associated error rates, this method can significantly contribute to the practical implementation of quantum algorithms on both current and future quantum processors. Future research could extend this method to work with more complex topologies and experimentally validate the proposed schemes.

\section{Acknowledgements}
We are grateful for useful discussion with Nicola Lo Gullo and the whole TORQS project members. 
The work is supported by Business Finland through the project "Towards Reliable Quantum Software Development: Approaches and Use Case" (TORQS), Grants No. 8582/31/2022 and 8436/31/2022,, and the project "Securing the Quantum Software Stack" (SeQuSoS), Grants No. 112/31/2024. M.S. acknowledges H2Future project through the Research Council of Finland (Grant. No. 352788) and the University of Oulu, and the Research Council of Finland Grant No. 316619.

\begin{thebibliography}{45}%
	\makeatletter
	\providecommand \@ifxundefined [1]{%
		\@ifx{#1\undefined}
	}%
	\providecommand \@ifnum [1]{%
		\ifnum #1\expandafter \@firstoftwo
		\else \expandafter \@secondoftwo
		\fi
	}%
	\providecommand \@ifx [1]{%
		\ifx #1\expandafter \@firstoftwo
		\else \expandafter \@secondoftwo
		\fi
	}%
	\providecommand \natexlab [1]{#1}%
	\providecommand \enquote  [1]{``#1''}%
	\providecommand \bibnamefont  [1]{#1}%
	\providecommand \bibfnamefont [1]{#1}%
	\providecommand \citenamefont [1]{#1}%
	\providecommand \href@noop [0]{\@secondoftwo}%
	\providecommand \href [0]{\begingroup \@sanitize@url \@href}%
	\providecommand \@href[1]{\@@startlink{#1}\@@href}%
	\providecommand \@@href[1]{\endgroup#1\@@endlink}%
	\providecommand \@sanitize@url [0]{\catcode `\\12\catcode `\$12\catcode `\&12\catcode `\#12\catcode `\^12\catcode `\_12\catcode `\%12\relax}%
	\providecommand \@@startlink[1]{}%
	\providecommand \@@endlink[0]{}%
	\providecommand \url  [0]{\begingroup\@sanitize@url \@url }%
	\providecommand \@url [1]{\endgroup\@href {#1}{\urlprefix }}%
	\providecommand \urlprefix  [0]{URL }%
	\providecommand \Eprint [0]{\href }%
	\providecommand \doibase [0]{https://doi.org/}%
	\providecommand \selectlanguage [0]{\@gobble}%
	\providecommand \bibinfo  [0]{\@secondoftwo}%
	\providecommand \bibfield  [0]{\@secondoftwo}%
	\providecommand \translation [1]{[#1]}%
	\providecommand \BibitemOpen [0]{}%
	\providecommand \bibitemStop [0]{}%
	\providecommand \bibitemNoStop [0]{.\EOS\space}%
	\providecommand \EOS [0]{\spacefactor3000\relax}%
	\providecommand \BibitemShut  [1]{\csname bibitem#1\endcsname}%
	\let\auto@bib@innerbib\@empty
	\bibitem [{\citenamefont {Moses}\ \emph {et~al.}(2023)\citenamefont {Moses}, \citenamefont {Baldwin}, \citenamefont {Allman}, \citenamefont {Ancona}, \citenamefont {Ascarrunz} \emph {et~al.}}]{Trappedion1}%
	\BibitemOpen
	\bibfield  {author} {\bibinfo {author} {\bibfnamefont {S.~A.}\ \bibnamefont {Moses}}, \bibinfo {author} {\bibfnamefont {C.~H.}\ \bibnamefont {Baldwin}}, \bibinfo {author} {\bibfnamefont {M.~S.}\ \bibnamefont {Allman}}, \bibinfo {author} {\bibfnamefont {R.}~\bibnamefont {Ancona}}, \bibinfo {author} {\bibfnamefont {L.}~\bibnamefont {Ascarrunz}}, \emph {et~al.},\ }\href {https://doi.org/10.1103/PhysRevX.13.041052} {\bibfield  {journal} {\bibinfo  {journal} {Phys. Rev. X}\ }\textbf {\bibinfo {volume} {13}},\ \bibinfo {pages} {041052} (\bibinfo {year} {2023})}\BibitemShut {NoStop}%
	\bibitem [{\citenamefont {Malinowski}\ \emph {et~al.}(2023)\citenamefont {Malinowski}, \citenamefont {Allcock},\ and\ \citenamefont {Ballance}}]{Trappedion2}%
	\BibitemOpen
	\bibfield  {author} {\bibinfo {author} {\bibfnamefont {M.}~\bibnamefont {Malinowski}}, \bibinfo {author} {\bibfnamefont {D.}~\bibnamefont {Allcock}},\ and\ \bibinfo {author} {\bibfnamefont {C.}~\bibnamefont {Ballance}},\ }\href {https://doi.org/10.1103/PRXQuantum.4.040313} {\bibfield  {journal} {\bibinfo  {journal} {PRX Quantum}\ }\textbf {\bibinfo {volume} {4}},\ \bibinfo {pages} {040313} (\bibinfo {year} {2023})}\BibitemShut {NoStop}%
	\bibitem [{\citenamefont {Bluvstein}\ \emph {et~al.}(2023)\citenamefont {Bluvstein}, \citenamefont {Evered}, \citenamefont {Geim}, \citenamefont {Li}, \citenamefont {Zhou} \emph {et~al.}}]{atom-array}%
	\BibitemOpen
	\bibfield  {author} {\bibinfo {author} {\bibfnamefont {D.}~\bibnamefont {Bluvstein}}, \bibinfo {author} {\bibfnamefont {S.~J.}\ \bibnamefont {Evered}}, \bibinfo {author} {\bibfnamefont {A.~A.}\ \bibnamefont {Geim}}, \bibinfo {author} {\bibfnamefont {S.~H.}\ \bibnamefont {Li}}, \bibinfo {author} {\bibfnamefont {H.}~\bibnamefont {Zhou}}, \emph {et~al.},\ }\href {https://api.semanticscholar.org/CorpusID:266052773} {\bibfield  {journal} {\bibinfo  {journal} {Nature}\ }\textbf {\bibinfo {volume} {626}},\ \bibinfo {pages} {58} (\bibinfo {year} {2023})}\BibitemShut {NoStop}%
	\bibitem [{\citenamefont {Kjaergaard}\ \emph {et~al.}(2020)\citenamefont {Kjaergaard}, \citenamefont {Schwartz}, \citenamefont {Braum\"{u}ller}, \citenamefont {Krantz}, \citenamefont {Wang} \emph {et~al.}}]{supercondu}%
	\BibitemOpen
	\bibfield  {author} {\bibinfo {author} {\bibfnamefont {M.}~\bibnamefont {Kjaergaard}}, \bibinfo {author} {\bibfnamefont {M.~E.}\ \bibnamefont {Schwartz}}, \bibinfo {author} {\bibfnamefont {J.}~\bibnamefont {Braum\"{u}ller}}, \bibinfo {author} {\bibfnamefont {P.}~\bibnamefont {Krantz}}, \bibinfo {author} {\bibfnamefont {J.~I.-J.}\ \bibnamefont {Wang}}, \emph {et~al.},\ }\href {https://doi.org/https://doi.org/10.1146/annurev-conmatphys-031119-050605} {\bibfield  {journal} {\bibinfo  {journal} {Annu. Rev. Condens. Matter Phys}\ }\textbf {\bibinfo {volume} {11}},\ \bibinfo {pages} {369} (\bibinfo {year} {2020})}\BibitemShut {NoStop}%
	\bibitem [{\citenamefont {Krantz}\ \emph {et~al.}(2019)\citenamefont {Krantz}, \citenamefont {Kjaergaard}, \citenamefont {Yan}, \citenamefont {Orlando}, \citenamefont {Gustavsson} \emph {et~al.}}]{supercondu1}%
	\BibitemOpen
	\bibfield  {author} {\bibinfo {author} {\bibfnamefont {P.}~\bibnamefont {Krantz}}, \bibinfo {author} {\bibfnamefont {M.}~\bibnamefont {Kjaergaard}}, \bibinfo {author} {\bibfnamefont {F.}~\bibnamefont {Yan}}, \bibinfo {author} {\bibfnamefont {T.~P.}\ \bibnamefont {Orlando}}, \bibinfo {author} {\bibfnamefont {S.}~\bibnamefont {Gustavsson}}, \emph {et~al.},\ }\href {https://doi.org/10.1063/1.5089550} {\bibfield  {journal} {\bibinfo  {journal} {Appl. Phys. Rev}\ }\textbf {\bibinfo {volume} {6}},\ \bibinfo {pages} {021318} (\bibinfo {year} {2019})}\BibitemShut {NoStop}%
	\bibitem [{\citenamefont {Cowtan}\ \emph {et~al.}(2019)\citenamefont {Cowtan}, \citenamefont {Dilkes}, \citenamefont {Duncan}, \citenamefont {Krajenbrink}, \citenamefont {Simmons},\ and\ \citenamefont {Sivarajah}}]{cowtan2019qubit}%
	\BibitemOpen
	\bibfield  {author} {\bibinfo {author} {\bibfnamefont {A.}~\bibnamefont {Cowtan}}, \bibinfo {author} {\bibfnamefont {S.}~\bibnamefont {Dilkes}}, \bibinfo {author} {\bibfnamefont {R.}~\bibnamefont {Duncan}}, \bibinfo {author} {\bibfnamefont {A.}~\bibnamefont {Krajenbrink}}, \bibinfo {author} {\bibfnamefont {W.}~\bibnamefont {Simmons}},\ and\ \bibinfo {author} {\bibfnamefont {S.}~\bibnamefont {Sivarajah}},\ }in\ \href {https://doi.org/10.4230/LIPIcs.TQC.2019.5} {\emph {\bibinfo {booktitle} {14th Conference on the Theory of Quantum Computation, Communication and Cryptography (TQC 2019)}}},\ Vol.\ \bibinfo {volume} {135}\ (\bibinfo {year} {2019})\ pp.\ \bibinfo {pages} {5:1--5:32}\BibitemShut {NoStop}%
	\bibitem [{\citenamefont {Bapat}\ \emph {et~al.}(2023)\citenamefont {Bapat}, \citenamefont {Childs}, \citenamefont {Gorshkov},\ and\ \citenamefont {Schoute}}]{PRXQuantumAdvantagesR}%
	\BibitemOpen
	\bibfield  {author} {\bibinfo {author} {\bibfnamefont {A.}~\bibnamefont {Bapat}}, \bibinfo {author} {\bibfnamefont {A.~M.}\ \bibnamefont {Childs}}, \bibinfo {author} {\bibfnamefont {A.~V.}\ \bibnamefont {Gorshkov}},\ and\ \bibinfo {author} {\bibfnamefont {E.}~\bibnamefont {Schoute}},\ }\href {https://doi.org/10.1103/PRXQuantum.4.010313} {\bibfield  {journal} {\bibinfo  {journal} {PRX Quantum}\ }\textbf {\bibinfo {volume} {4}},\ \bibinfo {pages} {010313} (\bibinfo {year} {2023})}\BibitemShut {NoStop}%
	\bibitem [{\citenamefont {Saeedi}\ \emph {et~al.}(2011)\citenamefont {Saeedi}, \citenamefont {Wille},\ and\ \citenamefont {Drechsler}}]{saeedi2011synthesis}%
	\BibitemOpen
	\bibfield  {author} {\bibinfo {author} {\bibfnamefont {M.}~\bibnamefont {Saeedi}}, \bibinfo {author} {\bibfnamefont {R.}~\bibnamefont {Wille}},\ and\ \bibinfo {author} {\bibfnamefont {R.}~\bibnamefont {Drechsler}},\ }\href {https://link.springer.com/article/10.1007/s11128-010-0201-2#citeas} {\bibfield  {journal} {\bibinfo  {journal} {Quantum Inf. Process.}\ }\textbf {\bibinfo {volume} {10}},\ \bibinfo {pages} {355} (\bibinfo {year} {2011})}\BibitemShut {NoStop}%
	\bibitem [{\citenamefont {Wille}\ \emph {et~al.}(2016)\citenamefont {Wille}, \citenamefont {Keszocze}, \citenamefont {Walter}, \citenamefont {Rohrs}, \citenamefont {Chattopadhyay},\ and\ \citenamefont {Drechsler}}]{shortestpath2}%
	\BibitemOpen
	\bibfield  {author} {\bibinfo {author} {\bibfnamefont {R.}~\bibnamefont {Wille}}, \bibinfo {author} {\bibfnamefont {O.}~\bibnamefont {Keszocze}}, \bibinfo {author} {\bibfnamefont {M.}~\bibnamefont {Walter}}, \bibinfo {author} {\bibfnamefont {P.}~\bibnamefont {Rohrs}}, \bibinfo {author} {\bibfnamefont {A.}~\bibnamefont {Chattopadhyay}},\ and\ \bibinfo {author} {\bibfnamefont {R.}~\bibnamefont {Drechsler}},\ }in\ \href {https://doi.org/10.1109/ASPDAC.2016.7428026} {\emph {\bibinfo {booktitle} {21st Asia and South Pacific Design Automation Conference (ASP-DAC)}}}\ (\bibinfo {year} {2016})\ pp.\ \bibinfo {pages} {292--297}\BibitemShut {NoStop}%
	\bibitem [{\citenamefont {Li}\ \emph {et~al.}(2019)\citenamefont {Li}, \citenamefont {Ding},\ and\ \citenamefont {Xie}}]{Sabre2019}%
	\BibitemOpen
	\bibfield  {author} {\bibinfo {author} {\bibfnamefont {G.}~\bibnamefont {Li}}, \bibinfo {author} {\bibfnamefont {Y.}~\bibnamefont {Ding}},\ and\ \bibinfo {author} {\bibfnamefont {Y.}~\bibnamefont {Xie}},\ }in\ \href {https://doi.org/10.1145/3297858.3304023} {\emph {\bibinfo {booktitle} {Proceedings of the Twenty-Fourth International Conference on Architectural Support for Programming Languages and Operating Systems}}},\ \bibinfo {series and number} {ASPLOS '19}\ (\bibinfo  {publisher} {Association for Computing Machinery},\ \bibinfo {address} {New York, NY, USA},\ \bibinfo {year} {2019})\ p.\ \bibinfo {pages} {1001–1014}\BibitemShut {NoStop}%
	\bibitem [{\citenamefont {Lao}\ \emph {et~al.}(2022)\citenamefont {Lao}, \citenamefont {van Someren}, \citenamefont {Ashraf},\ and\ \citenamefont {Almudever}}]{Lao2021}%
	\BibitemOpen
	\bibfield  {author} {\bibinfo {author} {\bibfnamefont {L.}~\bibnamefont {Lao}}, \bibinfo {author} {\bibfnamefont {H.}~\bibnamefont {van Someren}}, \bibinfo {author} {\bibfnamefont {I.}~\bibnamefont {Ashraf}},\ and\ \bibinfo {author} {\bibfnamefont {C.~G.}\ \bibnamefont {Almudever}},\ }\href {https://doi.org/10.1109/TCAD.2021.3057583} {\bibfield  {journal} {\bibinfo  {journal} {IEEE Transactions on Computer-Aided Design of Integrated Circuits and Systems}\ }\textbf {\bibinfo {volume} {41}},\ \bibinfo {pages} {359} (\bibinfo {year} {2022})}\BibitemShut {NoStop}%
	\bibitem [{\citenamefont {Li}\ \emph {et~al.}(2020)\citenamefont {Li}, \citenamefont {Meng}, \citenamefont {Zhang},\ and\ \citenamefont {Yu}}]{Li2020}%
	\BibitemOpen
	\bibfield  {author} {\bibinfo {author} {\bibfnamefont {Z.-T.}\ \bibnamefont {Li}}, \bibinfo {author} {\bibfnamefont {F.-X.}\ \bibnamefont {Meng}}, \bibinfo {author} {\bibfnamefont {Z.}~\bibnamefont {Zhang}},\ and\ \bibinfo {author} {\bibfnamefont {X.}~\bibnamefont {Yu}},\ }\href {https://link.springer.com/article/10.1007/s11128-020-02873-5} {\bibfield  {journal} {\bibinfo  {journal} {Quantum Inf. Process.}\ }\textbf {\bibinfo {volume} {19}} (\bibinfo {year} {2020})}\BibitemShut {NoStop}%
	\bibitem [{\citenamefont {Liu}\ \emph {et~al.}(2022)\citenamefont {Liu}, \citenamefont {Li},\ and\ \citenamefont {Zhou}}]{Liu2022}%
	\BibitemOpen
	\bibfield  {author} {\bibinfo {author} {\bibfnamefont {J.}~\bibnamefont {Liu}}, \bibinfo {author} {\bibfnamefont {P.}~\bibnamefont {Li}},\ and\ \bibinfo {author} {\bibfnamefont {H.}~\bibnamefont {Zhou}},\ }in\ \href {https://doi.org/10.1109/HPCA53966.2022.00058} {\emph {\bibinfo {booktitle} {2022 IEEE International Symposium on High-Performance Computer Architecture (HPCA)}}}\ (\bibinfo {year} {2022})\ pp.\ \bibinfo {pages} {709--725}\BibitemShut {NoStop}%
	\bibitem [{\citenamefont {Nishio}\ \emph {et~al.}(2020)\citenamefont {Nishio}, \citenamefont {Pan}, \citenamefont {Satoh}, \citenamefont {Amano},\ and\ \citenamefont {Meter}}]{Nishio2020}%
	\BibitemOpen
	\bibfield  {author} {\bibinfo {author} {\bibfnamefont {S.}~\bibnamefont {Nishio}}, \bibinfo {author} {\bibfnamefont {Y.}~\bibnamefont {Pan}}, \bibinfo {author} {\bibfnamefont {T.}~\bibnamefont {Satoh}}, \bibinfo {author} {\bibfnamefont {H.}~\bibnamefont {Amano}},\ and\ \bibinfo {author} {\bibfnamefont {R.~V.}\ \bibnamefont {Meter}},\ }\href {https://doi.org/10.1145/3386162} {\bibfield  {journal} {\bibinfo  {journal} {J. Emerg. Technol. Comput. Syst.}\ }\textbf {\bibinfo {volume} {16}},\ \bibinfo {pages} {25} (\bibinfo {year} {2020})}\BibitemShut {NoStop}%
	\bibitem [{\citenamefont {Chakrabarti}\ \emph {et~al.}(2011)\citenamefont {Chakrabarti}, \citenamefont {Sur-Kolay},\ and\ \citenamefont {Chaudhury}}]{chakrabarti2011linear}%
	\BibitemOpen
	\bibfield  {author} {\bibinfo {author} {\bibfnamefont {A.}~\bibnamefont {Chakrabarti}}, \bibinfo {author} {\bibfnamefont {S.}~\bibnamefont {Sur-Kolay}},\ and\ \bibinfo {author} {\bibfnamefont {A.}~\bibnamefont {Chaudhury}},\ }\href {https://arxiv.org/abs/1112.0564} {\bibfield  {journal} {\bibinfo  {journal} {arXiv:1112.0564}\ } (\bibinfo {year} {2011})}\BibitemShut {NoStop}%
	\bibitem [{\citenamefont {Shafaei}\ \emph {et~al.}(2013)\citenamefont {Shafaei}, \citenamefont {Saeedi},\ and\ \citenamefont {Pedram}}]{shafaei2013optimization}%
	\BibitemOpen
	\bibfield  {author} {\bibinfo {author} {\bibfnamefont {A.}~\bibnamefont {Shafaei}}, \bibinfo {author} {\bibfnamefont {M.}~\bibnamefont {Saeedi}},\ and\ \bibinfo {author} {\bibfnamefont {M.}~\bibnamefont {Pedram}},\ }in\ \href {https://doi.org/https://doi.org/10.1145/2463209.2488785} {\emph {\bibinfo {booktitle} {50th ACM/EDAC/IEEE Design Automation Conference (DAC)}}}\ (\bibinfo {year} {2013})\ pp.\ \bibinfo {pages} {1--6}\BibitemShut {NoStop}%
	\bibitem [{\citenamefont {Wille}\ \emph {et~al.}(2014)\citenamefont {Wille}, \citenamefont {Lye},\ and\ \citenamefont {Drechsler}}]{wille2014optimal}%
	\BibitemOpen
	\bibfield  {author} {\bibinfo {author} {\bibfnamefont {R.}~\bibnamefont {Wille}}, \bibinfo {author} {\bibfnamefont {A.}~\bibnamefont {Lye}},\ and\ \bibinfo {author} {\bibfnamefont {R.}~\bibnamefont {Drechsler}},\ }in\ \href {https://doi.org/10.1109/ASPDAC.2014.6742939} {\emph {\bibinfo {booktitle} {2014 19th Asia and South Pacific Design Automation Conference (ASP-DAC)}}}\ (\bibinfo {organization} {IEEE},\ \bibinfo {year} {2014})\ pp.\ \bibinfo {pages} {489--494}\BibitemShut {NoStop}%
	\bibitem [{\citenamefont {Lye}\ \emph {et~al.}(2015)\citenamefont {Lye}, \citenamefont {Wille},\ and\ \citenamefont {Drechsler}}]{lye2015determining}%
	\BibitemOpen
	\bibfield  {author} {\bibinfo {author} {\bibfnamefont {A.}~\bibnamefont {Lye}}, \bibinfo {author} {\bibfnamefont {R.}~\bibnamefont {Wille}},\ and\ \bibinfo {author} {\bibfnamefont {R.}~\bibnamefont {Drechsler}},\ }in\ \href {https://doi.org/10.1109/ASPDAC.2015.7059001} {\emph {\bibinfo {booktitle} {The 20th Asia and South Pacific Design Automation Conference}}}\ (\bibinfo {organization} {IEEE},\ \bibinfo {year} {2015})\ pp.\ \bibinfo {pages} {178--183}\BibitemShut {NoStop}%
	\bibitem [{\citenamefont {Bhattacharjee}\ and\ \citenamefont {Chattopadhyay}(2017)}]{bhattacharjee2017depth}%
	\BibitemOpen
	\bibfield  {author} {\bibinfo {author} {\bibfnamefont {D.}~\bibnamefont {Bhattacharjee}}\ and\ \bibinfo {author} {\bibfnamefont {A.}~\bibnamefont {Chattopadhyay}},\ }\href {https://doi.org/10.48550/arXiv.1703.08540} {\bibfield  {journal} {\bibinfo  {journal} {arXiv:1703.08540}\ } (\bibinfo {year} {2017})}\BibitemShut {NoStop}%
	\bibitem [{\citenamefont {Venturelli}\ \emph {et~al.}(2017)\citenamefont {Venturelli}, \citenamefont {Do}, \citenamefont {Rieffel},\ and\ \citenamefont {Frank}}]{venturelli2017temporal}%
	\BibitemOpen
	\bibfield  {author} {\bibinfo {author} {\bibfnamefont {D.}~\bibnamefont {Venturelli}}, \bibinfo {author} {\bibfnamefont {M.}~\bibnamefont {Do}}, \bibinfo {author} {\bibfnamefont {E.}~\bibnamefont {Rieffel}},\ and\ \bibinfo {author} {\bibfnamefont {J.}~\bibnamefont {Frank}},\ }in\ \href {https://dl.acm.org/doi/10.5555/3171837.3171907} {\emph {\bibinfo {booktitle} {Proceedings of the 26th International Joint Conference on Artificial Intelligence}}}\ (\bibinfo  {publisher} {AAAI Press},\ \bibinfo {year} {2017})\ p.\ \bibinfo {pages} {4440–4446}\BibitemShut {NoStop}%
	\bibitem [{\citenamefont {Venturelli}\ \emph {et~al.}(2018)\citenamefont {Venturelli}, \citenamefont {Do}, \citenamefont {Rieffel},\ and\ \citenamefont {Frank}}]{venturelli2018compiling}%
	\BibitemOpen
	\bibfield  {author} {\bibinfo {author} {\bibfnamefont {D.}~\bibnamefont {Venturelli}}, \bibinfo {author} {\bibfnamefont {M.}~\bibnamefont {Do}}, \bibinfo {author} {\bibfnamefont {E.}~\bibnamefont {Rieffel}},\ and\ \bibinfo {author} {\bibfnamefont {J.}~\bibnamefont {Frank}},\ }\href {https://iopscience.iop.org/article/10.1088/2058-9565/aaa331} {\bibfield  {journal} {\bibinfo  {journal} {Quantum Sci. Technol.}\ }\textbf {\bibinfo {volume} {3}},\ \bibinfo {pages} {025004} (\bibinfo {year} {2018})}\BibitemShut {NoStop}%
	\bibitem [{\citenamefont {Booth}\ \emph {et~al.}(2018)\citenamefont {Booth}, \citenamefont {Do}, \citenamefont {Beck}, \citenamefont {Rieffel}, \citenamefont {Venturelli} \emph {et~al.}}]{booth2018comparing}%
	\BibitemOpen
	\bibfield  {author} {\bibinfo {author} {\bibfnamefont {K.}~\bibnamefont {Booth}}, \bibinfo {author} {\bibfnamefont {M.}~\bibnamefont {Do}}, \bibinfo {author} {\bibfnamefont {J.}~\bibnamefont {Beck}}, \bibinfo {author} {\bibfnamefont {E.}~\bibnamefont {Rieffel}}, \bibinfo {author} {\bibfnamefont {D.}~\bibnamefont {Venturelli}}, \emph {et~al.},\ }in\ \href {https://doi.org/10.1609/icaps.v28i1.13920} {\emph {\bibinfo {booktitle} {Proceedings of the International Conference on Automated Planning and Scheduling}}},\ Vol.~\bibinfo {volume} {28}\ (\bibinfo {year} {2018})\ pp.\ \bibinfo {pages} {366--374}\BibitemShut {NoStop}%
	\bibitem [{\citenamefont {Oddi}\ and\ \citenamefont {Rasconi}(2018)}]{oddi2018greedy}%
	\BibitemOpen
	\bibfield  {author} {\bibinfo {author} {\bibfnamefont {A.}~\bibnamefont {Oddi}}\ and\ \bibinfo {author} {\bibfnamefont {R.}~\bibnamefont {Rasconi}},\ }in\ \href {https://doi.org/10.1007/978-3-319-93031-2_32} {\emph {\bibinfo {booktitle} {Integration of Constraint Programming, Artificial Intelligence, and Operations Research: 15th International Conference, CPAIOR 2018, Delft, The Netherlands, June 26--29, 2018, Proceedings 15}}}\ (\bibinfo {organization} {Springer},\ \bibinfo {year} {2018})\ pp.\ \bibinfo {pages} {446--461}\BibitemShut {NoStop}%
	\bibitem [{\citenamefont {Tang}\ \emph {et~al.}(2024)\citenamefont {Tang}, \citenamefont {Duan}, \citenamefont {Kharkov}, \citenamefont {Fakoor}, \citenamefont {Kessler},\ and\ \citenamefont {Shi}}]{tang2024alpharouter}%
	\BibitemOpen
	\bibfield  {author} {\bibinfo {author} {\bibfnamefont {W.}~\bibnamefont {Tang}}, \bibinfo {author} {\bibfnamefont {Y.}~\bibnamefont {Duan}}, \bibinfo {author} {\bibfnamefont {Y.}~\bibnamefont {Kharkov}}, \bibinfo {author} {\bibfnamefont {R.}~\bibnamefont {Fakoor}}, \bibinfo {author} {\bibfnamefont {E.}~\bibnamefont {Kessler}},\ and\ \bibinfo {author} {\bibfnamefont {Y.}~\bibnamefont {Shi}},\ }in\ \href {https://doi.org/10.1109/qce60285.2024.00112} {\emph {\bibinfo {booktitle} {2024 IEEE International Conference on Quantum Computing and Engineering (QCE)}}}\ (\bibinfo  {publisher} {IEEE},\ \bibinfo {year} {2024})\ p.\ \bibinfo {pages} {930–940}\BibitemShut {NoStop}%
	\bibitem [{\citenamefont {Kremer}\ \emph {et~al.}(2024)\citenamefont {Kremer}, \citenamefont {Villar}, \citenamefont {Paik}, \citenamefont {Duran}, \citenamefont {Faro},\ and\ \citenamefont {Cruz-Benito}}]{RL_based_David}%
	\BibitemOpen
	\bibfield  {author} {\bibinfo {author} {\bibfnamefont {D.}~\bibnamefont {Kremer}}, \bibinfo {author} {\bibfnamefont {V.}~\bibnamefont {Villar}}, \bibinfo {author} {\bibfnamefont {H.}~\bibnamefont {Paik}}, \bibinfo {author} {\bibfnamefont {I.}~\bibnamefont {Duran}}, \bibinfo {author} {\bibfnamefont {I.}~\bibnamefont {Faro}},\ and\ \bibinfo {author} {\bibfnamefont {J.}~\bibnamefont {Cruz-Benito}},\ }\href {https://arxiv.org/abs/2405.13196} {\bibfield  {journal} {\bibinfo  {journal} {arXiv:2405.13196}\ } (\bibinfo {year} {2024})}\BibitemShut {NoStop}%
	\bibitem [{\citenamefont {Sivarajah}\ \emph {et~al.}(2020)\citenamefont {Sivarajah}, \citenamefont {Dilkes}, \citenamefont {Cowtan}, \citenamefont {Simmons}, \citenamefont {Edgington} \emph {et~al.}}]{Sivarajah2020}%
	\BibitemOpen
	\bibfield  {author} {\bibinfo {author} {\bibfnamefont {S.}~\bibnamefont {Sivarajah}}, \bibinfo {author} {\bibfnamefont {S.}~\bibnamefont {Dilkes}}, \bibinfo {author} {\bibfnamefont {A.}~\bibnamefont {Cowtan}}, \bibinfo {author} {\bibfnamefont {W.}~\bibnamefont {Simmons}}, \bibinfo {author} {\bibfnamefont {A.}~\bibnamefont {Edgington}}, \emph {et~al.},\ }\href {http://dx.doi.org/10.1088/2058-9565/ab8e92} {\bibfield  {journal} {\bibinfo  {journal} {Quantum Sci. Technol.}\ }\textbf {\bibinfo {volume} {6}},\ \bibinfo {pages} {014003} (\bibinfo {year} {2020})}\BibitemShut {NoStop}%
	\bibitem [{\citenamefont {Bennett}\ \emph {et~al.}(1993)\citenamefont {Bennett}, \citenamefont {Brassard}, \citenamefont {Cr\'{e}peau}, \citenamefont {Jozsa}, \citenamefont {Peres} \emph {et~al.}}]{Bennett1993}%
	\BibitemOpen
	\bibfield  {author} {\bibinfo {author} {\bibfnamefont {C.~H.}\ \bibnamefont {Bennett}}, \bibinfo {author} {\bibfnamefont {G.}~\bibnamefont {Brassard}}, \bibinfo {author} {\bibfnamefont {C.}~\bibnamefont {Cr\'{e}peau}}, \bibinfo {author} {\bibfnamefont {R.}~\bibnamefont {Jozsa}}, \bibinfo {author} {\bibfnamefont {A.}~\bibnamefont {Peres}}, \emph {et~al.},\ }\href {https://doi.org/10.1103/physrevlett.70.1895} {\bibfield  {journal} {\bibinfo  {journal} {Phys. Rev. Lett.}\ }\textbf {\bibinfo {volume} {70}},\ \bibinfo {pages} {1895–1899} (\bibinfo {year} {1993})}\BibitemShut {NoStop}%
	\bibitem [{\citenamefont {Gottesman}\ and\ \citenamefont {Chuang}(1999)}]{Gottesman1999}%
	\BibitemOpen
	\bibfield  {author} {\bibinfo {author} {\bibfnamefont {D.}~\bibnamefont {Gottesman}}\ and\ \bibinfo {author} {\bibfnamefont {I.~L.}\ \bibnamefont {Chuang}},\ }\href {https://doi.org/10.1038/46503} {\bibfield  {journal} {\bibinfo  {journal} {Nature}\ }\textbf {\bibinfo {volume} {402}},\ \bibinfo {pages} {390} (\bibinfo {year} {1999})}\BibitemShut {NoStop}%
	\bibitem [{\citenamefont {Gisin}\ and\ \citenamefont {Thew}(2007)}]{Gisin2007}%
	\BibitemOpen
	\bibfield  {author} {\bibinfo {author} {\bibfnamefont {N.}~\bibnamefont {Gisin}}\ and\ \bibinfo {author} {\bibfnamefont {R.}~\bibnamefont {Thew}},\ }\href {https://doi.org/10.1038/nphoton.2007.22} {\bibfield  {journal} {\bibinfo  {journal} {Nat. Photon.}\ }\textbf {\bibinfo {volume} {1}},\ \bibinfo {pages} {165} (\bibinfo {year} {2007})}\BibitemShut {NoStop}%
	\bibitem [{\citenamefont {Briegel}\ \emph {et~al.}(1998)\citenamefont {Briegel}, \citenamefont {D\"{u}r}, \citenamefont {Cirac},\ and\ \citenamefont {Zoller}}]{Briegel1998}%
	\BibitemOpen
	\bibfield  {author} {\bibinfo {author} {\bibfnamefont {H.-J.}\ \bibnamefont {Briegel}}, \bibinfo {author} {\bibfnamefont {W.}~\bibnamefont {D\"{u}r}}, \bibinfo {author} {\bibfnamefont {J.~I.}\ \bibnamefont {Cirac}},\ and\ \bibinfo {author} {\bibfnamefont {P.}~\bibnamefont {Zoller}},\ }\href {https://doi.org/10.1103/physrevlett.81.5932} {\bibfield  {journal} {\bibinfo  {journal} {Phys. Rev. Lett.}\ }\textbf {\bibinfo {volume} {81}},\ \bibinfo {pages} {5932} (\bibinfo {year} {1998})}\BibitemShut {NoStop}%
	\bibitem [{\citenamefont {Gordon}\ and\ \citenamefont {Rigolin}(2010)}]{Gordon2010}%
	\BibitemOpen
	\bibfield  {author} {\bibinfo {author} {\bibfnamefont {G.}~\bibnamefont {Gordon}}\ and\ \bibinfo {author} {\bibfnamefont {G.}~\bibnamefont {Rigolin}},\ }\href {https://doi.org/10.1016/j.optcom.2009.09.028} {\bibfield  {journal} {\bibinfo  {journal} {Opt. Commun.}\ }\textbf {\bibinfo {volume} {283}},\ \bibinfo {pages} {184} (\bibinfo {year} {2010})}\BibitemShut {NoStop}%
	\bibitem [{\citenamefont {Raussendorf}\ and\ \citenamefont {Briegel}(2001)}]{Raussendorf2001}%
	\BibitemOpen
	\bibfield  {author} {\bibinfo {author} {\bibfnamefont {R.}~\bibnamefont {Raussendorf}}\ and\ \bibinfo {author} {\bibfnamefont {H.~J.}\ \bibnamefont {Briegel}},\ }\href {https://doi.org/10.1103/physrevlett.86.5188} {\bibfield  {journal} {\bibinfo  {journal} {Phys. Rev. Lett.}\ }\textbf {\bibinfo {volume} {86}},\ \bibinfo {pages} {5188} (\bibinfo {year} {2001})}\BibitemShut {NoStop}%
	\bibitem [{\citenamefont {B\"aumer}\ \emph {et~al.}(2024{\natexlab{a}})\citenamefont {B\"aumer}, \citenamefont {Tripathi}, \citenamefont {Wang}, \citenamefont {Rall}, \citenamefont {Chen}, \citenamefont {Majumder}, \citenamefont {Seif},\ and\ \citenamefont {Minev}}]{Bumer2024}%
	\BibitemOpen
	\bibfield  {author} {\bibinfo {author} {\bibfnamefont {E.}~\bibnamefont {B\"aumer}}, \bibinfo {author} {\bibfnamefont {V.}~\bibnamefont {Tripathi}}, \bibinfo {author} {\bibfnamefont {D.~S.}\ \bibnamefont {Wang}}, \bibinfo {author} {\bibfnamefont {P.}~\bibnamefont {Rall}}, \bibinfo {author} {\bibfnamefont {E.~H.}\ \bibnamefont {Chen}}, \bibinfo {author} {\bibfnamefont {S.}~\bibnamefont {Majumder}}, \bibinfo {author} {\bibfnamefont {A.}~\bibnamefont {Seif}},\ and\ \bibinfo {author} {\bibfnamefont {Z.~K.}\ \bibnamefont {Minev}},\ }\href {https://doi.org/10.1103/PRXQuantum.5.030339} {\bibfield  {journal} {\bibinfo  {journal} {PRX Quantum}\ }\textbf {\bibinfo {volume} {5}},\ \bibinfo {pages} {030339} (\bibinfo {year} {2024}{\natexlab{a}})}\BibitemShut {NoStop}%
	\bibitem [{\citenamefont {B\"aumer}\ \emph {et~al.}(2024{\natexlab{b}})\citenamefont {B\"aumer}, \citenamefont {Tripathi}, \citenamefont {Seif}, \citenamefont {Lidar},\ and\ \citenamefont {Wang}}]{Baumer2024_2}%
	\BibitemOpen
	\bibfield  {author} {\bibinfo {author} {\bibfnamefont {E.}~\bibnamefont {B\"aumer}}, \bibinfo {author} {\bibfnamefont {V.}~\bibnamefont {Tripathi}}, \bibinfo {author} {\bibfnamefont {A.}~\bibnamefont {Seif}}, \bibinfo {author} {\bibfnamefont {D.}~\bibnamefont {Lidar}},\ and\ \bibinfo {author} {\bibfnamefont {D.~S.}\ \bibnamefont {Wang}},\ }\href {https://doi.org/10.1103/PhysRevLett.133.150602} {\bibfield  {journal} {\bibinfo  {journal} {Phys. Rev. Lett.}\ }\textbf {\bibinfo {volume} {133}},\ \bibinfo {pages} {150602} (\bibinfo {year} {2024}{\natexlab{b}})}\BibitemShut {NoStop}%
	\bibitem [{\citenamefont {Devulapalli}\ \emph {et~al.}(2024)\citenamefont {Devulapalli}, \citenamefont {Schoute}, \citenamefont {Bapat}, \citenamefont {Childs},\ and\ \citenamefont {Gorshkov}}]{Devulapalli2024}%
	\BibitemOpen
	\bibfield  {author} {\bibinfo {author} {\bibfnamefont {D.}~\bibnamefont {Devulapalli}}, \bibinfo {author} {\bibfnamefont {E.}~\bibnamefont {Schoute}}, \bibinfo {author} {\bibfnamefont {A.}~\bibnamefont {Bapat}}, \bibinfo {author} {\bibfnamefont {A.~M.}\ \bibnamefont {Childs}},\ and\ \bibinfo {author} {\bibfnamefont {A.~V.}\ \bibnamefont {Gorshkov}},\ }\href {https://doi.org/10.1103/physrevresearch.6.033313} {\bibfield  {journal} {\bibinfo  {journal} {Phys. Rev. Research}\ }\textbf {\bibinfo {volume} {6}},\ \bibinfo {pages} {033313} (\bibinfo {year} {2024})}\BibitemShut {NoStop}%
	\bibitem [{\citenamefont {Hillmich}\ \emph {et~al.}(2021)\citenamefont {Hillmich}, \citenamefont {Zulehner},\ and\ \citenamefont {Wille}}]{Hillmich2021}%
	\BibitemOpen
	\bibfield  {author} {\bibinfo {author} {\bibfnamefont {S.}~\bibnamefont {Hillmich}}, \bibinfo {author} {\bibfnamefont {A.}~\bibnamefont {Zulehner}},\ and\ \bibinfo {author} {\bibfnamefont {R.}~\bibnamefont {Wille}},\ }in\ \href {https://doi.org/10.1145/3394885.3431604} {\emph {\bibinfo {booktitle} {Proceedings of the 26th Asia and South Pacific Design Automation Conference}}},\ \bibinfo {series} {ASPDAC '21}, Vol.\ \bibinfo {volume} {1508}\ (\bibinfo  {publisher} {ACM},\ \bibinfo {year} {2021})\ p.\ \bibinfo {pages} {792–797}\BibitemShut {NoStop}%
	\bibitem [{\citenamefont {Zulehner}\ \emph {et~al.}(2019)\citenamefont {Zulehner}, \citenamefont {Paler},\ and\ \citenamefont {Wille}}]{Zulehner2019}%
	\BibitemOpen
	\bibfield  {author} {\bibinfo {author} {\bibfnamefont {A.}~\bibnamefont {Zulehner}}, \bibinfo {author} {\bibfnamefont {A.}~\bibnamefont {Paler}},\ and\ \bibinfo {author} {\bibfnamefont {R.}~\bibnamefont {Wille}},\ }\href {https://doi.org/10.1109/tcad.2018.2846658} {\bibfield  {journal} {\bibinfo  {journal} {IEEE Trans. Comput.-Aided Des. Integr. Circuits Syst.}\ }\textbf {\bibinfo {volume} {38}},\ \bibinfo {pages} {1226} (\bibinfo {year} {2019})}\BibitemShut {NoStop}%
	\bibitem [{\citenamefont {Padda}\ \emph {et~al.}(2024)\citenamefont {Padda}, \citenamefont {Tham}, \citenamefont {Brodutch},\ and\ \citenamefont {Touchette}}]{Padda2024}%
	\BibitemOpen
	\bibfield  {author} {\bibinfo {author} {\bibfnamefont {G.}~\bibnamefont {Padda}}, \bibinfo {author} {\bibfnamefont {E.}~\bibnamefont {Tham}}, \bibinfo {author} {\bibfnamefont {A.}~\bibnamefont {Brodutch}},\ and\ \bibinfo {author} {\bibfnamefont {D.}~\bibnamefont {Touchette}},\ }in\ \href {https://doi.org/10.1109/qce60285.2024.00206} {\emph {\bibinfo {booktitle} {2024 IEEE International Conference on Quantum Computing and Engineering (QCE)}}}\ (\bibinfo  {publisher} {IEEE},\ \bibinfo {year} {2024})\ p.\ \bibinfo {pages} {1770–1776}\BibitemShut {NoStop}%
	\bibitem [{\citenamefont {Eisert}\ \emph {et~al.}(2000)\citenamefont {Eisert}, \citenamefont {Jacobs}, \citenamefont {Papadopoulos},\ and\ \citenamefont {Plenio}}]{Eisert2000}%
	\BibitemOpen
	\bibfield  {author} {\bibinfo {author} {\bibfnamefont {J.}~\bibnamefont {Eisert}}, \bibinfo {author} {\bibfnamefont {K.}~\bibnamefont {Jacobs}}, \bibinfo {author} {\bibfnamefont {P.}~\bibnamefont {Papadopoulos}},\ and\ \bibinfo {author} {\bibfnamefont {M.~B.}\ \bibnamefont {Plenio}},\ }\href {https://doi.org/10.1103/PhysRevA.62.052317} {\bibfield  {journal} {\bibinfo  {journal} {Phys. Rev. A}\ }\textbf {\bibinfo {volume} {62}},\ \bibinfo {pages} {052317} (\bibinfo {year} {2000})}\BibitemShut {NoStop}%
	\bibitem [{\citenamefont {Aseguinolaza}\ \emph {et~al.}(2023)\citenamefont {Aseguinolaza}, \citenamefont {Sobrino}, \citenamefont {Sobrino}, \citenamefont {Jornet-Somoza},\ and\ \citenamefont {Borge}}]{errorestimation}%
	\BibitemOpen
	\bibfield  {author} {\bibinfo {author} {\bibfnamefont {U.}~\bibnamefont {Aseguinolaza}}, \bibinfo {author} {\bibfnamefont {N.}~\bibnamefont {Sobrino}}, \bibinfo {author} {\bibfnamefont {G.}~\bibnamefont {Sobrino}}, \bibinfo {author} {\bibfnamefont {J.}~\bibnamefont {Jornet-Somoza}},\ and\ \bibinfo {author} {\bibfnamefont {J.}~\bibnamefont {Borge}},\ }\href {https://api.semanticscholar.org/CorpusID:259309445} {\bibfield  {journal} {\bibinfo  {journal} {Quantum Inf. Process.}\ }\textbf {\bibinfo {volume} {23}},\ \bibinfo {pages} {181} (\bibinfo {year} {2023})}\BibitemShut {NoStop}%
	\bibitem [{\citenamefont {Quetschlich}\ \emph {et~al.}(2023)\citenamefont {Quetschlich}, \citenamefont {Burgholzer},\ and\ \citenamefont {Wille}}]{quetschlich2023mqtbench}%
	\BibitemOpen
	\bibfield  {author} {\bibinfo {author} {\bibfnamefont {N.}~\bibnamefont {Quetschlich}}, \bibinfo {author} {\bibfnamefont {L.}~\bibnamefont {Burgholzer}},\ and\ \bibinfo {author} {\bibfnamefont {R.}~\bibnamefont {Wille}},\ }\href {https://doi.org/10.22331/q-2023-07-20-1062} {\bibfield  {journal} {\bibinfo  {journal} {{Quantum}}\ }\textbf {\bibinfo {volume} {6}},\ \bibinfo {pages} {1062} (\bibinfo {year} {2023})},\ \bibinfo {note} {{{MQT Bench}} is available at \url{https://www.cda.cit.tum.de/mqtbench/}}\BibitemShut {NoStop}%
	\bibitem [{\citenamefont {IBMQuantum}()}]{IBMQuantum}%
	\BibitemOpen
	\bibfield  {author} {\bibinfo {author} {\bibnamefont {IBMQuantum}},\ }\href {https://quantum.ibm.com/} {\bibinfo {title} {https://quantum.ibm.com/}}\BibitemShut {NoStop}%
	\bibitem [{\citenamefont {Buhrman}\ \emph {et~al.}(2024)\citenamefont {Buhrman}, \citenamefont {Folkertsma}, \citenamefont {Loff},\ and\ \citenamefont {Neumann}}]{Buhrman2024statepreparation}%
	\BibitemOpen
	\bibfield  {author} {\bibinfo {author} {\bibfnamefont {H.}~\bibnamefont {Buhrman}}, \bibinfo {author} {\bibfnamefont {M.}~\bibnamefont {Folkertsma}}, \bibinfo {author} {\bibfnamefont {B.}~\bibnamefont {Loff}},\ and\ \bibinfo {author} {\bibfnamefont {N.~M.~P.}\ \bibnamefont {Neumann}},\ }\href {https://doi.org/10.22331/q-2024-12-09-1552} {\bibfield  {journal} {\bibinfo  {journal} {{Quantum}}\ }\textbf {\bibinfo {volume} {8}},\ \bibinfo {pages} {1552} (\bibinfo {year} {2024})}\BibitemShut {NoStop}%
	\bibitem [{\citenamefont {Yeo}\ \emph {et~al.}(2025)\citenamefont {Yeo}, \citenamefont {Kim}, \citenamefont {Sohn},\ and\ \citenamefont {Jeong}}]{yeo2025reducing}%
	\BibitemOpen
	\bibfield  {author} {\bibinfo {author} {\bibfnamefont {H.}~\bibnamefont {Yeo}}, \bibinfo {author} {\bibfnamefont {H.~E.}\ \bibnamefont {Kim}}, \bibinfo {author} {\bibfnamefont {I.}~\bibnamefont {Sohn}},\ and\ \bibinfo {author} {\bibfnamefont {K.}~\bibnamefont {Jeong}},\ }\href {https://doi.org/10.48550/arXiv.2501.02929} {\bibfield  {journal} {\bibinfo  {journal} {arXiv:2501.02929}\ } (\bibinfo {year} {2025})}\BibitemShut {NoStop}%
	\bibitem [{\citenamefont {Cai}\ \emph {et~al.}(2023)\citenamefont {Cai}, \citenamefont {Babbush}, \citenamefont {Benjamin}, \citenamefont {Endo}, \citenamefont {Huggins}, \citenamefont {Li}, \citenamefont {McClean},\ and\ \citenamefont {O’Brien}}]{Cai23}%
	\BibitemOpen
	\bibfield  {author} {\bibinfo {author} {\bibfnamefont {Z.}~\bibnamefont {Cai}}, \bibinfo {author} {\bibfnamefont {R.}~\bibnamefont {Babbush}}, \bibinfo {author} {\bibfnamefont {S.~C.}\ \bibnamefont {Benjamin}}, \bibinfo {author} {\bibfnamefont {S.}~\bibnamefont {Endo}}, \bibinfo {author} {\bibfnamefont {W.~J.}\ \bibnamefont {Huggins}}, \bibinfo {author} {\bibfnamefont {Y.}~\bibnamefont {Li}}, \bibinfo {author} {\bibfnamefont {J.~R.}\ \bibnamefont {McClean}},\ and\ \bibinfo {author} {\bibfnamefont {T.~E.}\ \bibnamefont {O’Brien}},\ }\href {https://doi.org/10.1103/RevModPhys.95.045005} {\bibfield  {journal} {\bibinfo  {journal} {Rev. Mod. Phys.}\ }\textbf {\bibinfo {volume} {95}},\ \bibinfo {pages} {045005} (\bibinfo {year} {2023})}\BibitemShut {NoStop}%
\end{thebibliography}
%
\end{document}